\documentclass{article}
\usepackage{axodraw}
\newcommand{\beq}{\begin{equation}}
\newcommand{\eeq}{\end{equation}}
\newcommand{\beqa}{\begin{eqnarray}}
\newcommand{\eeqa}{\end{eqnarray}}
\newcommand{\bear}{\begin{array}}
\newcommand{\ear}{\end{array}}

\newcommand{\pth}{perturbation theory}

\newcommand{\GB}{Gupta--Bleuler}
\newcommand{\lb}[1]{\label{#1}}
\newcommand{\Ref}[1]{(\ref{#1})}
\newcommand{\sla}[1]{{#1}\!\!\!/}

\newcommand{\der}{\partial}

\newcommand{\psar}{\bar{\psi}}
\newcommand{\Psar}{\bar{\Psi}}

\newcommand{\Omg}{\Omega}

\newcommand{\gl}{\!\!\!\!&=&\!\!\!\!}

\newcommand{\Tilps}{\tilde{\Psar}}
\newcommand{\tilp}{\tilde{\psi}}
\newcommand{\Tilp}{\tilde{\Psi}}
\newcommand{\half}{\frac{1}{2}}

\newcommand{\nonu}{\nonumber \\}
\newcommand{\tilf}{\tilde{f}}

\begin{document}

\title{What is the Magnetic Moment of the Electron?}
\author{Othmar Steinmann\\Fakult\"at f\"ur Physik\\Universit\"at Bielefeld\\D-33501 Bielefeld,
Germany}
\date{}
\maketitle

\begin{abstract}
Because of infrared  effects the charged sectors of QED contain no
eigenstates of the mass operator. The electron is therefore not
definable as a Wigner particle. There exists no sharp,
unambiguous, definition of the notion of a 1-electron state. The
assignment of a fixed value of the magnetic moment -- or similar
quantities -- to the electron is therefore at first problematic.
It is not clear a priori that such a notion is meaningful.
Conventionally this problem is solved by first calculating the
desired quantity in an IR-regularized theory and then removing the
regularization. If this method yields a finite value, that is
considered sufficient proof of its soundness. This is clearly less
than satisfactory. Here we propose a more convincing way of
defining the intrinsic magnetic moment of the electron, which does
not use any regularizations and is not based on an interaction
with external fields. A pseudostatic 1-electron state is defined
in a phenomenological way. Its magnetic moment, as defined here,
does not depend on the unavoidable ambiguities inherent in this
definition. The method leads to the same analytic expression as
the conventional approach, thus preserving the excellent agreement
between theory and experiment.
\end{abstract}

\section{Introduction}
The spectacular accuracy with which the theoretical QED--values of
the anomalous  magnetic moments of the electron and the muon agree
with measurement\footnote{A fairly recent review is \cite{HK}.
The latest measurement in the
muon case is  reported in \cite{g2}.}
is one of the major triumphs of relativistic quantum field theory.
It must be said, however, that from the point of view of a
rigorous formulation of field theory, the theoretical derivation
of these numbers leaves much to be desired. The method typically
used, as briefly described in \cite{Kin}, can be epitomized  by
the following quote from this reference: ``The magnetic property
of an electron can be studied   most conveniently by examining its
scattering by a static magnetic field". But the standard
scattering  formalism used in carrying out this program is not
really applicable to QED even in the absence of an external field,
due to the notorious infrared (IR) problems. Taking these problems
seriously it is found that for charged particles a 1-particle
state is a much more complex object than usually assumed. In
particular it is {\em not} definable as an eigenstate of the mass
operator $M^2=P_{\mu}P^{\mu}$ ($P_{\mu}$ the 4-momentum operator).
Green's functions and the like cannot be meaningfully restricted
to the mass shell. A ``1-particle state" can therefore not be
specified by a 3-dimensional wave function. Customarily this fact
is described by stating that a charged particle is necessarily
accompanied by a cloud of soft photons, the exact composition of
this cloud not being derivable from first principles. What is
fixed is, crudely speaking, only the form of the singularity of
$n(\omega)$ for $\omega\to0$, when $n(\omega)$ is the number  of
soft photons of energy $\omega$.\footnote{This description uses
the language of the interaction representation, which is
mathematically unsound because of Haag's theorem \cite{StW,BLOT},
a fact that is unfortunately  still largely ignored in the
literature.} Besides invalidating the conventional scattering
formalism, this unavoidable vagueness of the 1-particle states
creates an obvious problem with the definition of  their magnetic
moment and similar quantities. Might they not be indeterminate
because depending on the shape of the photon cloud?

 The problem becomes even more serious if the system is acted upon
  by an external electromagnetic field. This destroys the
 Poincar\'e invariance of the theory, in particular the
 translation invariance which is a powerful tool in the ordinary
 treatment. This raises, for instance, the important question of
 how to define the vacuum state, which seems to be a prerequisite
 for a meaningful definition of a 1-particle state. Also, it is
 not clear that the quantum fields can still be expected to
 satisfy asymptotic conditions like the LSZ condition, that is, to
 converge in a suitable sense to free fields for large positive or
 negative times, unless the external fields tend to zero fast
 enough in this limit, which is clearly not the case for static
 external fields.

   As a result, the scattering amplitude underlying the
   conventional formalism does actually not exist. In a
   perturbative treatment this non-existence  manifests itself most
   prominently by the IR divergences of the formal expression of
   the amplitude. The traditional way of handling this problem
   consists in starting from an IR-regularized theory, typically
   by introducing a positive photon mass $\mu$, and letting $\mu $
tend to zero in the final expression for whatever physical
quantity one is interested in. But  the fact that this derivation
yields a finite (i.e. divergence-free) value of the magnetic
moment does not make it any less suspect. It has hardly more than
a heuristic value.  Indeed, it has been shown in \cite{bk} that in
the case of scattering cross sections this method very likely
produces erroneous results.

In view of the undeniable success in describing observation, the
theoretical formula for the magnetic moment thus deviously
obtained is, however, undoubtedly correct. But because of the
importance of this result a more convincing derivation is
desirable. Such a derivation will be proposed in the present
paper.

\section{Outline of the Method}

The method to be presented is based on the particle notion
introduced in \cite{bk}\footnote{This reference will henceforth be
quoted as BK.}  Particles play no fundamental role in the theory.
They are secondary phenomenological objects which are useful for
the description of observations. We are especially interested in
the magnetic moment of the electron due to its spin. The major
ingredients of the formalism are the notion of an approximate
1-electron state and an intrinsic definition of the magnetic
moment not relying on its response to external fields.

We work throughout in the Heisenberg picture, using \pth, since an
exact treatment is beyond the possibilities of present-day field
theory. An ``approximate electron state'' is defined by (see
 Eq.\ (14.36) of BK)
 \beq
\Phi_f = \Psi(f)\,\Omg, \qquad \Psi(f) = \int dx\,f(x)\,\Psi(x) =
\int dp\,\tilf (p)\,\tilde{\Psi}(p) .                   \lb{1}
 \eeq
Here $\Omg$ is the vacuum state and $f(x)$ is a test spinor, that
is a 4-component wave function whose components are infinitely
differentiable functions of $x$ with strong decrease at infinity.
Its Fourier transform
 \beq
\tilf(p) = (2\pi)^{-3/2} \int  dx\,e^{-ipx}\,f(x) \lb{2}
 \eeq
has a small compact support centered at a point $P$ on the
negative mass shell $\{P^2=m^2, P_0<0\}$, $m$ the mass of the
electron. $\Psi$ is the electron field in a physical gauge, which
we choose for convenience to be a rotation invariant gauge like,
for instance, the Coulomb gauge. $\tilf(p)$ shall be smooth in the
sense that it is $C^{\infty}$ and slowly varying, i.e. not
containing any violent wiggles. $f\,\Psi$ is an abbreviation for
$f_{\rho}\Psi_{\rho}$ summed over the four spinor indices
$\rho$.\footnote{For spinor indices we use the summation
convention without regard to their position, while for Minkowski
indices we sum over indices occurring twice only if one stands
downstairs, one upstairs.} For later purposes we note
 \beq
\big(\Psi(f)\big)^* = \int dx\, \Psar(x)\,\gamma^0\,f^*(x) = \int
dp\  \tilde{\Psar}(-p)\,\gamma^0\,\tilf^*(p) . \lb{3}
 \eeq

In BK it has been shown that $\Phi_f$ exhibits the behavior of a
free particle of mass $m$  if monitored by detectors placed at
macroscopic spacetime distances from one another. This explains
the appellation ``approximate electron state".

  \medskip

The operator of the magnetic moment we define by taking over the
corresponding expression from classical magnetostatics (see
\cite{Jak}, Sect.\ 5.6)
 \beq
\mathbf{M} = \half \int d^3x\,\big({\bf x}\times {\bf j}({\bf
x})\big) , \lb{4}
 \eeq
where $j(x)$ is the operator         of the electromagnetic
current density. The magnetic moment of a stationary state $\Phi$
we define as the expectation value
 \beq
{\bf m} = \frac{(\Phi,\,{\bf M}\,\Phi)}{(\Phi,\Phi)}      \lb{5}
 \eeq
of ${\bf M}$ in $\Phi$. This formula is not immediately applicable
to our problem because $\Phi_f$ is clearly not stationary. But we
can make $\Phi_f$ almost stationary by concentrating the support
of $\tilf$ around the zero-momentum point $P=(-m,{\bf 0})$ and
making the diameter of this support arbitrarily small.
Unfortunately it is not meaningful to let this diameter shrink to
zero. The states $\Phi_f$ would not converge in this limit. And
even if we considered only expectation values like in Eq. \Ref{5},
the shrinking of the $\tilf$-support would lead to an increasing
delocalization of the state in $x$-space which might interfere
with the ${\bf x}$-integration in \Ref{4}. This integration might
not commute with the $\tilf$-limit, with awkward consequences.

We will therefore use a weaker notion of a ``static limit". All
our calculations will be carried out in \pth, writing
$(\Phi_f,\,{\bf M}\,\Phi_f)$ and $(\Phi_f,\Phi_f)$ as sums over
generalized Feynman graphs (see below). In their integrands in
$p$-space we replace $p$ by $P$ in all slowly varying factors. The
resulting expression we call the ``static limit" of the graph by
abuse of language. Note that the wave function $\tilf$ itself is {\em not}
slowly varying despite its smoothness, because its assumed tiny
support necessitates large variations over small
distances. Another point to be noted is that we are only
interested in the contribution of the spin to the magnetic moment,
excluding the effects of the orbital motion. We therefore consider
only s-states in which an orbital contribution is not to be
expected. This means that we assume the wave function $\tilf$,
more exactly each of its four components $\tilf_{\rho}(p)$, to be
invariant under space rotations: $\tilf_{\rho}(p)$ is assumed to
depend only on the two variables $p^0$ and $|{\bf p}|^2$:
 \beq
\tilf_{\rho}(p) = \tilf_{\rho}\big(p^0,\,|{\bf p}|^2\big)\  .
\lb{6} \eeq

The desired intrinsic magnetic moment of the electron is then
defined as the static limit of
 \beq
{\bf m} = \frac{(\Phi_f,\,{\bf M}\,\Phi_f)}{(\Phi_f,\Phi_f)} \ .
\lb{7}        \eeq

Rotational invariance being assumed, we can restrict ourselves to
considering the 3-component
 \beq
m_3 = \frac{(\Phi_f,\,M_3\,\Phi_f)}{(\Phi_f,\Phi_f)}\ .      \lb{8}
 \eeq

The field $\Psi(x)$ transforms under the rotation $R$ as
 \beq
\Psi(Rx) = S(R)\,U(R)\,\Psi(x)\,U^*(R)\ .            \lb{9}
 \eeq
$U(R)$ is the unitary representation of the rotation group defined
on the state space of the theory, $S(R)$ is the well-known
4-dimensional spinor representation acting on the spinor index
$\rho$ of $\Psi_{\rho}$. For $R$ a rotation through the angle
$\chi$ around the 3-axis we define the spin matrix $\Sigma_3$ and
the operator $J_3$ of angular momentum by
 \beq
S(R) = e^{i\chi\Sigma_3}\ ,\qquad U(R) = e^{-i\chi J_3}\ . \lb{10}
 \eeq
With the conventions\footnote{The Dirac matrices are
$\gamma^0=\left(\bear{cc} 0&{\bf 1} \\ {\bf 1}&0 \ear \right),\
\gamma^i=\left(\bear{cc} 0&-\sigma_i \\ \sigma_i&0 \ear \right)\
,\ i=1,2,3$. All entries in these matrices are themselves
$2\times2$matrices. The $\sigma_i$ are the Pauli matrices.} of BK
we find
 \beq
\Sigma_3 =
-\,\half\left(\bear{cccc}1&0&0&0\\0&-1&0&0\\0&0&1&0\\0&0&0&-1\ear\right)
= \half\,\sigma^{21}                  \lb{11}
 \eeq
with
 \beq
\sigma^{\mu\nu} :=
\frac{i}{2}\,\big[\gamma^{\mu},\gamma^{\nu}\big]\ .        \lb{12}
 \eeq
$J_3$ may be decomposed into an orbital part $L_3$ and a spin part
$S_3$:
 \beq
J_3 = L_3 + S_3\ ,                           \lb{13}
 \eeq
with $L_3$ the 3-component of the standard expression ${\bf
L}={\bf x}\times{\bf P}\:,{\bf P}=-i\nabla$.  Defining $\Phi(x) =
\Psi(x)\Omg$, ${\bf L}\Omg={\bf S}\Omg=0$, and using Eq.~\Ref{9}
for an infinitesimal angle $\chi$ we find
 \beq
S_3\Phi(x) = \Sigma_3\Phi(x)\             \lb{14}
 \eeq
which relation extends by linearity to $\Phi_f$. We define the
spin content of $\Phi_f$ as
 \beq
\frac{(\Phi_f,S_3\Phi_f)}{(\Phi_f,\Phi_f)}         \lb{15}
 \eeq
taken in the static limit. The ``gyromagnetic ratio" $g$ of the
electron is defined by
 \beq
(\Phi_f,\,M_3\,\Phi_f) =
-\,\frac{eg}{2m}\,\left(\Phi_f,\,S_3\,\Phi_f\right)\ ,   \lb{16}
 \eeq
both sides being taken in the static limit. The coupling constant
$e$ is defined to be the elementary charge unit, i.e. the charge
of the positron, not that of the electron as often done in the
literature. This explains the negative sign in the right-hand side
of Eq.~\Ref{16}. $e/2m$ is known as the Bohr magneton. The
existence of such a constant $g$, solving Eq.~\Ref{16} irrespective
of the exact form of the chosen wave function, is by no means
obvious. But the two sides of \Ref{16}  clearly transform in the
same way under rotations around the 3-axis. Hence we can choose
$f$ to be an eigenfunction of $\Sigma_3$ with eigenvalue $\pm1/2$
without restriction of generality. We have then $S_3\Phi_f =
\pm\,\half\,\Phi_f$.

  \medskip

The problem that we want to solve is, then, to prove that
Eq.~\Ref{16} can be satisfied in \pth \  for a suitable choice of
$g$, to show how this $g$ can be calculated as a formal power
series
 \beq
g = \sum^{\infty}_{\sigma=0} g_{\sigma} e^{\sigma}\ , \lb{17}
 \eeq
and to show that the result thus obtained coincides with the
conventional expression used in the well-known numerical
evaluations of $g$.

More concretely we propose to determine the perturbative
coefficients $g_{\sigma}$ as follows. The expectation values of
$M_3$ and $S_3$ occurring in Eq.~\Ref{16} can be expanded in
perturbation series with the methods developed in BK, as will be
explained in Sect.\ 4. Assume that $g_{\tau}$ is known for all
$\tau<\sigma$. Then we find $g_{\sigma}$ as solution of
 \beq
 \frac{g_{\sigma}}{2m}\,(\Phi_f,\,S_3\,\Phi_f)_0
 = - (\Phi_f,\,M_3\,\Phi_f)_{\sigma+1} - \frac{1}{2m}
 \sum_{\tau=0}^{\sigma-1}
 g_{\tau}(\Phi_f,\,S_3\,\Phi_f)_{\sigma-\tau}\ .  \lb{18}
   \eeq
Here $(\Phi_f,\,\mathcal{O}\,\Phi_f)_{\tau}$ denotes the
coefficient of order $\tau$ in the perturbative expansion of
$(\Phi_f,\mathcal{O}\,\Phi_f)$ taken in the static limit. The main
problem here, solved in Sect. 4, is the proof that this leads to an
$f$-independent $g_{\sigma}$. That the result coincides with the
conventional one will be shown in Sect.\ 5.

 \section{Calculation of $g_o$}
In the lowest order $\sigma=0$ of \pth\ there are no radiative
corrections and therefore no IR problems. Moreover, $\Psi$ is the
local, covariant, free Dirac field $\psi(x)$. Hence the
conventional method is quite unobjectionable and trustworthy. The
fact that our method yields the same result may help to create
some confidence in its credibility.

In our method $g_o$ is determined from
 \beq
L := \frac{g_o}{2m}\,(\Phi_f,\,S_3\,\Phi_f)_o =
-(\Phi_f,\,M_3\,\Phi_f)_1 =: R\ .               \lb{19}
 \eeq
We may assume that $S_3\Phi_f = \half\,\Phi_f$. For $L$ we find
then
 \beqa
L \gl - \frac{g_o}{4m} \int
dq\,\delta_-(q)\,\tilf(q)\,(\sla{q}+m)\,\gamma^0 \tilf^*(q) \nonu
 \gl \frac{g_o}{4m} \int \frac{d^3q}{2\omega} \tilf(-\omega,\,{\bf
 q})\,(\omega\gamma^0-q_i \gamma^i-m)\,\gamma^0
 \tilf^*(-\omega,\,{\bf q})                        \lb{20}
  \eeqa
with $\delta_-(q) = \theta(-q_0)\,\delta(q^2-m^2)$,
$\omega=\omega({\bf q})=\omega_{{\bf q}}=\sqrt{{\bf q}^2+m^2}$. In
the static limit the term $q_i\gamma^i$ vanishes and $\omega\to
m$, so that in this limit we have
 \beq
L = \frac{g_o}{8m} \int d^3q\,\tilf(-\omega,\,{\mathbf
q})\,(\gamma^0-1)\,\gamma^0\tilf^*(-\omega,\,{\mathbf q})\ .
\lb{21}      \eeq

The right-hand side $R$ of Eq.\Ref{19} is\footnote{We exhibit
spinor indices explicitly when the order of multiplication does
not correspond to the order shown in the equation.}
 \beqa
R \gl -\,\half \int
dx\,dy\,d^3u\,f_{\alpha}(x)\,\big(\gamma^0f^*(y)\big)_{\beta}
\nonu  & & \hspace{-1cm}\times
\Big(\Omg,\,\psar_{\beta}(y)\,\big[
u^1\!:\!\psar(u)\,\gamma^2\psi(u)\!: -
u^2\!:\!\psar(u)\,\gamma^1\psi(u)\!:\big]\,\psi_{\alpha}(x)\,\Omg\Big)\Big|_{u^0=0}.
 \lb{22}  \eeqa
 The value $u^0=0$ is chosen for convenience. In fact, $R$ does
 not depend on $u^0$ in the static limit. $\psi$ being a free
 Dirac field we can calculate the vacuum expectation values in
 this expression with the help of Wick's theorem. Fourier
 transforms are defined as
  \beqa
 \psi(x) \gl (2\pi)^{-3/2} \int dp\,e^{-ipx}\,\tilp(p)\ , \nonu
  f(x) \gl (2\pi)^{-5/2} \int dq \,e^{iqx}\,\tilf(q)\ . \lb{23}
 \eeqa
  We obtain the $p$-space form of $R$:
 \beqa
  R \gl -\,\frac{i}{2} \int
  dp\,\delta_-(p)\,dq\,\delta_-(q)\,dk_0\,\tilf(q)\,(\sla{q}+m)\,
  \Big[\gamma^2\frac{\der}{\der k_1}  - \gamma^1\frac{\der}{\der k_2}\Big]
  \delta^4(k+q-p)  \nonu
  & & \times\: (\sla{p}+m)\,\gamma^0\tilf^*(p)\Big|_{\mathbf{k}=0}
  \nonu
  \gl -\,\frac{i}{8} \int \frac{d^3q}{\omega({\bf p})}\,\frac{d^3q}{\omega({\bf
  q})}\,\tilf(-\omega({\bf q}),{\bf q})\,\big(\omega({\bf
  q})\gamma^0-q_i\gamma^i-m\big)\ \times   \nonu
  & & \hspace*{-8mm} \Big[\gamma^2\,\frac{\der}{\der k_1} - \gamma^1\,
  \frac{\der}{\der
  k_2}\Big]\,\delta^3(\mathbf{k+q-p})\, \big(\omega({\bf
  p})\,\gamma^0-p_i\gamma^i-m\big)\,\gamma^0 \tilf^*(-\omega({\bf
  p}),\,{\bf p})\Big|_{\mathbf{k}=0}   \ .  \nonu
   & &   \lb{24}    \eeqa
Because of the derivations in the integrand we cannot yet neglect
the weakly varying ${\bf p}$--${\bf q}$--terms. A simple algebraic
calculation shows that
 \beqa
 & &
 \delta_-(q)\,\delta_-(p)\,(\sla{q}+m)\,\gamma^{\mu}(\sla{p}+m)
  \nonu
 \gl
 \delta_-(q)\,\delta_-(p)\,(\sla{q}+m)\,\left(-\,\frac{1}{4m}\,[\sla{p}-\sla{q},
 \gamma^{\mu}] + \frac{1}{2m}\,(p^{\mu}+q^{\mu})\right)
 \,(\sla{p}+m)\,.   \lb{25}
\eeqa
 This is a corollary to the Gordon decomposition of the current
 operator into a spin part and an orbital part.  Let us consider
 the contribution of the $p^{\mu}$-term to $R$. It contains the
 expression
  \[\left(p^2\frac{\der}{\der k_1} - p^1\frac{\der}{\der k_2}\right)\!
   \delta^3(\mathbf{k+q-p})\bigg|_{{\bf k}=0}  = -\left(p^2\frac{\der}{\der p_1} +
   p^1\frac{\der}{\der p_2}\right)\! \delta^3(\mathbf{k+q-p})\bigg|_{{\bf k}=0}. \]
Since the derivations act no longer on $k$ we can now set
$\mathbf{k}=0$  in the $\delta^3$-factor. Moreover,
\[ -p^2\frac{\der}{\der p_1} + p^1\frac{\der}{\der p_2} =
p_2\frac{\der}{\der p_1} - p_1\frac{\der}{\der p_2} = -iL_3\ , \]
 with $L_3 $ the generator of geometrical rotations around the
 $x$-axis. $L_3$ acting on $\delta^3(\mathbf{p-q})$ can be
 transferred to the other $p$-dependent factors in $R$ through
 integration by parts. But $L_3$  annihilates the rotation
 invariant factors $\delta_-(p),\,\omega({\bf p})$, and
 $\tilf^*(p)$. The only remaining term contains $L_3\sla{p} = i(p^2\gamma^1 -
 p^1\gamma^2)$, which vanishes in the static limit. In the same
 way the irrelevance of the $q^{\mu}$-term in Eq.~\Ref{25} is
 shown.

 There remains the commutator in \Ref{25} to be discussed.
 Consider the $\gamma^2$-term in the last term of Eq.~\Ref{24}.  We
 replace
  \[ \gamma^2\frac{\der}{\der k_1}\,\delta^3(\mathbf{k+q-p})\bigg|_{{\bf
  k}=0}  \]
by
 \[ -\,\frac{1}{4m}\,\big[\big(\omega({\bf q})-\omega({\bf p})\big)\,
 \gamma^0 + (p_i-q_i)\,\gamma^i\,,\,\gamma^2\big]\,\frac{\der}{\der q_1}
 \,\delta^3(\mathbf{q-p})\ . \]
Integration by parts transfers the derivation to the other
$q$-dependent factors  which are all $C^{\infty}$. The derivative
of the factors other than $[\cdots,\cdots]$ gets multiplied with
 \[ \Big[\big(\omega({\bf q})-\omega({\bf p})\big)\gamma^0  +
 (p_i-q_i)\,\gamma^i\,,\,\gamma^2\Big]\,\delta^3({\bf q-p}) = 0\ .   \]
The derivative of the commutator is
 \[ -\,\left[\frac{q_1}{\omega({\bf q})}\,\gamma^0 - \gamma^1\, ,\,\gamma^2\right]\ .
  \]
The $\gamma^0$-term vanishes in the static limit and we remain
with $[\gamma^1,\gamma^2]$. In the same way the $\gamma^1$-term in
\Ref{24}  yields a factor $-[\gamma^2,\gamma^1] =
[\gamma^1,\gamma^2]$. The result is, taking again the static limit
\[ R = -\,\frac{1}{4m} \int d^3q\,\tilf\big(-\omega({\bf q}),{\bf q}\big)
\,(\gamma^0-1)\,\Sigma_3\,(\gamma^0-1)\,\gamma^0\,\tilf^*\big(-\omega({\bf
q}),{\bf q}\big)\ .
\]
Using that $\Sigma_3$ commutes with $\gamma^0$ and that
$\Sigma_3\tilf^* = \half\,\tilf^*$ by assumption, we obtain
 \beq
R = \frac{1}{4m} \int d^3q\,\tilf(-\omega,{\bf
q})\,(\gamma^0-1)\,\gamma^0\, \tilf^*(-\omega,{\bf q})\ . \lb{26}
 \eeq
Equating this with $L$ as given in Eq.\Ref{21} we find
 \beq     g_o = 2\ ,   \lb{27}   \eeq
the classical Dirac result.

\section{Perturbation Theory in General Order}

The coefficient $g_{\sigma}$ of order  $e^{\sigma}$ of the
gyromagnetic ratio is determined by Eq.~\Ref{18}. It is our task to
show that this equation indeed possesses a unique solution which
does not depend on the choice of $f$, within the restrictions
specified in Sects. 2 and 3. In particular, we assume again that
$f$ is an eigenfunction of $\Sigma_3$  with eigenvalue 1/2. The
left-hand side of \Ref{18} is then given by the expression
\Ref{21} with $g_o$ replaced by $g_{\sigma}$. The right-hand side
contains the $g_{\tau}$ with $\tau<\sigma$ which we assume to be
known. The expectation values $(\Phi_f,\,M_3\,\Phi_f)_{\sigma+1}$
and $(\Phi_f,\,S_3\Phi_f)_{\sigma-\tau} =
\half\,(\Phi_f,\,\Phi_f)_{\sigma-\tau}$ can be calculated by the
methods explained in BK and stated more explicitly below. They
contain the Wightman functions
$\big(\Omg,\tilde{\bar{\Psi}}(-p)\,\tilde{\Psi}(q)\,\Omg\big)$
or
$\big(\Omg,\tilde{\bar{\Psi}}(-p)\,\tilde{j^{\mu}}(k)\,\tilde{\Psi}(q)\,\Omg\big)$
respectively. Their perturbative expressions are given in
unrenormalized form in Sect. 9.3 of BK\footnote{The sign rule iv) on p.119 contains
an embarrassing error. It should state that {\em each} fermionic
cross line pointing from a higher to a lower sector contributes a
factor -1, not only those in closed loops.} for
the Gupta-Bleuler fields, amended for physical gauges in Sect. 12.3.


\begin{figure}[ht]
\begin{center}
\begin{picture}(250,95)
\SetOffset(0,-10)
\SetWidth{0.8} \DashArrowArcn(85,38)(41,135,45){3}
\DashArrowArcn(165,38)(41,135,45){3}
 \DashArrowArcn(125,-13)(113,135,45){3}
\SetWidth{0.5}
 \GOval(45,67)(15,25)(0){0.6}
\GOval(125,67)(15,25)(0){0.6} \GOval(205,67)(15,25)(0){0.6}
 \SetWidth{0.8}
\ArrowLine(20,67)(0,67)      \Text(10,70)[b]{$p$}
 \Line(20,67)(70,67)
  \ArrowLine(100,67)(70,67)      \Text(85,64)[t]{$r$}
\Line(100,67)(150,67)
 \ArrowLine(180,67)(150,67)  \Text(165,64)[t]{$s$}
\Line(180,67)(230,67)
 \ArrowLine(250,67)(230,67)      \Text(240,70)[b]{$q$}
 \SetWidth{0.3}
\Line(80,105)(80,40)   \Line(170,105)(170,40)
 \Text(40,46)[t]{\large $S_1^-$}   \Text(125,46)[t]{\large $S^+_2$}
 \Text(210,46)[t]{\large $S^-_3$}
 \Text(125,10)[b]{{\bf Fig.~1.} \small A S-graph}

\end{picture}
\end{center}
\end{figure}

Take first the ``S-terms" containing
 \[ (\Phi_f,\Phi_f)_{\rho} = \int dp\,dq\,\tilf(q)\,
\big(\Omg,\tilde{\bar{\Psi}}(-p)\,\tilde{\Psi}(q)\,\Omega\big)_{\!\rho}\,\tilf^*(p)\
.  \]
 The vacuum expectation value in this expression can be written as
 a sum over 3-sector graphs of the general form shown in Fig.1.
 The external sectors $S_1$ and $S_3$ are chosen to be
 $T^-$-sectors, the internal sector $S_2$, which may be empty, a
 $T^+$-sector. The bubbles denote subgraphs. There may be any number of
 photon cross lines. The sets of photon
 momenta of cross lines connecting $S_1$ with $S_2$, $S_2$ with
 $S_3$, $S_1$ with $S_3$, respectively, are denoted by $L_{\ell},\ L_r,
 \ L_m$, the number of elements of these sets by $|L_{\ell}|,\ |L_r|,\ |L_m|$.
 The external fermion lines are connected by an unbroken
 fermion trajectory.
 Because of the assumed small support of $\tilf$
 there cannot be any fermionic cross lines other than those of the trajectory.
  The graph rules inside the bubbles are the
 ordinary Feynman rules in a $T^+$-sector, their antichronological
 form in a $T^-$-sector.
 This differs from the chronological form
 by sign changes of all vertex factors and propagators and of the
 $i\varepsilon$-prescription. The lines crossing sector boundaries
 carry propagators containing factors $\delta_-(r) = \theta(-r_o)\,
 \delta(r^2-m^2)$, analogously for $s$, for the trajectory lines,
 and $\delta_+(\ell) = \theta(\ell_o)\,\delta(\ell^2)$
   for photon lines. The physical external vertices  introduced in
   Sect.\ 12.3 of BK can only occur at the beginning and the end of
   the trajectory, that is in the $T^-$-sectors, while the rules
   in the internal sector $S_2$  are the ordinary \GB\ ones. The
   individual graphs are in general IR divergent. But these
   divergences cancel in the sums over all graphs of order $\rho$,
   as has been shown in Chap.\ 11 of BK.

In the sequel it is occasionally convenient to consider our
expressions as limits of the corresponding expressions in
``massive QED", in which the photon propagators are regularized
with the help of a small photon mass $\mu>0$. This means that
$(k^2\pm i\varepsilon)^{-1}\,,\,\delta(k^2)$, are replaced by
$(k^2-\mu^2\pm i\varepsilon)^{-1},\,\delta(k^2-\mu^2)$. This renders
the individual graphs IR convergent. The limit  $\mu\to0$ does in
general not exist for individual graphs. But it does exist for the
sum over all graphs of a given order and yields the correct
expression. This allows to discard graphs which vanish identically
for $\mu>0$, as is the case for our graphs if
$|L_{\ell}|>0,\;|L_r|=|L_m|=0$, or $|L_r|>0,\;|L_{\ell}|=|L_m|=0$,
because e.g. $\delta^m_-(r)\prod_{\ell_i\in L_{\ell}}
\delta^{\mu}_+(\ell_i)\, \delta^m_-(s=r-\sum\ell_i)\equiv0$ for
$\mu>0$. In the surviving graphs we will in general set $\mu=0$
directly. Of course, the limit $\mu\to0$ must be taken {\em
before} any other limit, especially before the static limit.

UV divergent bubbles inside the sectors are renormalized by
subtraction at vanishing external momenta. Fermionic self-energy
parts (SEPs) are then caused to vanish at the mass shell by an
additional {\em finite} mass renormalization.  But no additional
field renormalization is effected for giving the residue of the
1-electron pole   a desired value, since the 1-particle
singularity of the clothed electron propagator is not a pole due
to IR effects.

If $|L_{\ell}|+|L_m|=0$ or $|L_r|+|L_m|=0$, there is a factor
$\delta_-(q)$ from a trajectory cross line present, so that only
the restriction
 \beq
  \tilf_M({\bf q}) :=\tilf\big(-\omega({\bf q}),\,{\bf q}\big)
  \lb{29}     \eeq
of $\tilf(q)$ to the mass shell contributes,  in accordance with
the desired result. If neither of these conditions is satisfied,
then the off-mass-shell values of $\tilf$ contribute to the
individual graphs. Accordingly we separate the set of graphs of
order $\rho$ into two classes, the $\tilf_M-\tilf_M$ class and the
$\tilf-\tilf$ class. Unfortunately, the IR cancellations
mentioned above  involve graphs from both classes, they are not
operative within a class. But we will find that instead the IR
divergences inside a class are cancelled by divergences of the
corresponding class of the ``M-terms" in Eq.\Ref{18}, the terms
containing $M_3$.


\begin{figure}[ht]
\begin{center}
\begin{picture}(250,95)
\SetOffset(0,-10)
\SetWidth{0.8} \DashArrowArcn(85,38)(41,135,45){3}
\DashArrowArcn(165,38)(41,135,45){3}
 \DashArrowArcn(125,-13)(113,135,45){3}
\SetWidth{0.5}
 \GOval(45,67)(15,25)(0){0.6}
\GOval(125,67)(15,25)(0){0.6} \GOval(205,67)(15,25)(0){0.6}
 \SetWidth{0.8}
\ArrowLine(20,67)(0,67)      \Text(10,70)[b]{$p$}
 \Line(20,67)(70,67)
  \ArrowLine(100,67)(70,67)      \Text(85,64)[t]{$r$}
\Line(100,67)(150,67)
 \ArrowLine(180,67)(150,67)  \Text(165,64)[t]{$s$}
\Line(180,67)(230,67)
 \ArrowLine(250,67)(230,67)      \Text(240,70)[b]{$q$}
 \SetWidth{0.3}
\Line(80,105)(80,40)   \Line(170,105)(170,40)
 \SetWidth{0.5}
 \Line(125,82)(123,85) \Line(125,82)(127,85)
 \Line(123,85)(127,85)  \Text(125,87)[b]{$k$}
 \Text(40,46)[t]{\large $S_1^-$}   \Text(125,46)[t]{\large $S^+_2$}
 \Text(210,46)[t]{\large $S^-_3$}
 \Text(125,10)[b]{{\bf Fig.\ 2.} \small A M-graph}

\end{picture}
\end{center}
\end{figure}

 \medskip

Let us turn to these M-terms. They involve the vacuum
expectation value
$\big(\Omega,\,\tilde{\bar{\Psi}}(-p)\,\tilde{j^{\mu}}(k)\,\tilde{\Psi}(q)\,
\Omg\big)_{\sigma+1}$ which can again be expressed as a sum over
3-sector graphs of order $\sigma+1$, but this time with all
sectors being external. They are of the general form shown in Fig.~2.
 The only difference to the S-graphs of Fig.~1 is that now the
 central  $S_2$-bubble  contains an external $j^{\mu}$-vertex with
 one amputated line (see Fig.~3). Its vertex factor is
   $(2\pi)^{-3/2}\gamma^{\mu}\delta^4(k+s'-r')$. Inserting this into
 the definition \Ref{4} of ${\bf M}$, which yields
 \[ M_3 = \half \int d^3x\,\big(x^1j^2(x) - x^2j^1(x)\big)\big|_{x^o=0}\ , \]
and Fourier transforming, we find that
$\big(\Omega,\,\Tilps(-p)\,M_3\,\Tilp(q)\,\Omg\big)_{\sigma+1}$ is
given by a sum over the same graphs, only with the $j$-vertex
replaced by an $M$-vertex
carrying the vertex factor
 \beqa   & &
-\,\frac{i}{2} \int dk_o\,\left(\gamma^{\mu}\frac{\der}{\der k_1}
- \gamma^1\,\frac{\der}{\der
k_2}\right)\delta^4(k+s'-r')\bigg|_{{\bf k}=0}    \nonu
 \gl -\,\frac{i}{2}\,\left(\gamma^2\,\frac{\der}{\der k_1} -
 \gamma^1\,\frac{\der}{\der
 k_2}\right)\delta^3(\mathbf{k+s'-r'})\bigg|_{{\bf k}=0}    \ .  \lb{28}
 \eeqa
 Here $k$ is the original $\tilde{j}$-variable.  Henceforth the
 name ``M-graphs" will be used for graphs containing this
 vertex factor, while those with the original j-vertex are
 called ``j-graphs".


\begin{figure}[ht]
 \begin{center}
 \begin{picture}(155,16)
 \SetWidth{0.8}
\ArrowLine(28,2)(0,2)            \Text(14,5)[b]{$r'$}
\ArrowLine(56,2)(28,2)            \Text(42,5)[b]{$s'$}
 \SetWidth{0.5}
 \Line(28,2.3)(26,5.3) \Line(28,2.3)(30,5.3)  \Line(26,5.3)(30,5.3)
  \Text(28,8)[b]{$k$}
 \Text(155,1)[br]{{\bf Fig.~3.} \small A j-vertex}

 \end{picture}
 \end{center}
\end{figure}

In contrast to the similar looking S-graphs, the external
variables $p,\,q$, are now independent. So, if $|L_r|+|L_m|=0$,
then $q$ is restricted to the mass shell but $p$ is not. Hence
$\tilf(q)$ contributes only via its mass-shell restriction
$\tilf_M$, while $\tilf^*(p)$ also contributes off mass-shell
unless we have $L_{\ell}=0$ too. But if $|L_{\ell}| = |L_r| =
|L_m| = 0$, then only the combination $\tilf_M$--$\tilf_M^*$
contributes, as is the case for the left-hand side of Eq.~\Ref{18}.
In order for \Ref{18} to be meaningful, the graphs on its
right-hand side containing the combinations $\tilf$--$\tilf^*$,
$\tilf$--$\tilf_M^*$, $\tilf_M$--$\tilf^*$, must add to zero.  To
establish this is our foremost task. This hinges on the ability to
handle the IR structure of our expressions.

At first let us disregard the singularities inside the sectors,
that is  we assume that the subgraphs shown as bubbles in Figs. 1
and 2 are smooth functions of their external variables
$r,\,s,\,k$, and the photon cross momenta $\{\ell_i\}$.   The
$\ell_j$ are negligibly small in the static limit on account of
the small $\tilf$-support and momentum conservation.
$r=p+\sum_{L_{\ell}\cup L_m}\ell_j$ and $s = q+\sum_{L_r\cup
L_m}\ell_j$ must lie on the negative mass shell, $\ell_j$  on the
positive mass shell, $p$ and $q$ in the support of $\tilf$, which
is only possible for small $\ell_j$.

Let $\Delta_p =p_o+m$ and $\Delta_q = q_o+m$ measure the distance
of $p$ and $q$ from the mass shell.  If $|L_m|=0$ and $|L_r|>0$
there occurs a phase space integral
 \beq
  I_r = \int \prod_{L_r}\,\frac{d^3\ell_i}{2|{\bf \ell_i}|}\,
  \frac{\delta(s_o+\omega_s)}{2\omega_s}   \lb{30}
 \eeq
with $s=q+\sum_{L_r}\ell_i$. But $\omega_{{\bf s}} =
\omega(\mathbf{q+\sum\ell_i}) =
m+\mathcal{O}\big((\mathbf{q+\sum\ell_i})^2\big)$, and this may be
replaced by $m$ in the static limit even if occurring inside a
singular factor. Hence $\delta(s_o+\omega_s)\sim
\delta(\Delta_q+\sum|\mathcal{\ell}_i|)$. For $\Delta_q\to0$ ,
hence $|\mathcal{\ell}_i|\to0$, the integral is easily seen to
vanish like
 \beq
  I_r(\Delta_q) = \mathcal{O}\!\left(|\Delta_q|^{2|L_r|-1}\right)\ .
  \lb{31}
 \eeq
 In the same way one finds a factor
  \beq
   I_{\ell}(\Delta_p) =
   \mathcal{O}\!\left(|\Delta_p|^{2L_{\ell}-1}\right)   \lb{32}
  \eeq
if $|L_{\ell}|>0,\;|L_m|=0$. For arbitrary $|L_m|$ we obtain the
general result, defining $\Delta = \big(\Delta_p{}^2 +
\Delta_q{}^2\big)^{1/2}$:
 \beq
  I_m(\Delta) = \int\prod\frac{d^3\ell_i}{2|\mathbf{\ell}_i|}\,
  \frac{\delta(s_o+\omega_{\mathbf{s}})}{2\omega_{\mathbf{s}}}\,
  \frac{\delta(r_o+\omega_{\mathbf{r}})}{2\omega_{\mathbf{r}}} =
    \mathcal{O}\!\left(|\Delta|^{2(|L_{\ell}|+|L_r|+|L_m|-1)}\right)\ .
    \lb{33}        \eeq
This simple consideration is not correct  if $|L_{\ell}|=|L_r|=0$.
But this case will not be needed later on.

 \medskip

The vanishing of the factors $I$ in the static limit is offset  by
the singular behavior of the other factors. The left-most sector
$S_1$ in Figs.\ 1, 2, contains the propagator $(p^2-m^2)^{-1}$ which
diverges like $\Delta^{-1}$ in the static limit, which implies
$\Delta\to0$.\footnote{In fact $(p^2-m^2)^{-1}$  diverges like
$\Delta_p{}^{-1}$ even if $\Delta_q\neq0$  and therefore
$\Delta\neq0$. But we are only concerned with the total order of
the singularity occurring in the static limit, in which both
$\Delta_p$ and $\Delta_q$ tend to zero. Future statements about
the order of $(\Delta\to0)$ singularities must also be understood
in this sense.} If no bubble is present in $S_1$, then this is
replaced by $\delta^m_-(p)$ which is singular of order
$\Delta^{-1}$ too. The same holds for the right-most sector $S_3$.
The bubbles in all three sectors may be singular themselves,
contrary to our provisional assumption. Singularities of power
type occur in bubbles which are 1-particle reducible (1PI), meaning that
they can be partitioned into two disconnected parts by cutting a
single line. If this line is a photon line, then the bubble
contributes the factor $B_1\,(L,L)^{-1}\,B_2$, where
$L=\sum'\ell_i$ is a sum over a subset $\sigma'$ of $\{\ell_i\}$,
$B_1$ is a subgraph with external variables $\sigma'$ and $L$, and
$B_2$ with external variables $\sigma\backslash\sigma'$ and $L$,
if $\sigma=\{p,q,\ell_i\}$. Inserting this new singular factor
into the phase integrals \Ref{30} and \Ref{33} we obtain at first
an additional factor $\Delta^{-2}$ in the estimated order of
vanishing for $\Delta\to0$. However, this is offset by the fact
that the residue factor $B_1$, depending exclusively on the
photonic variables $\ell_i$, vanishes at least like $\Delta^2$ in
the static limit (which implies $\ell_i\to0$) due to the
Ward-Takahashi identity and to covariance. Hence these
singularities do not disturb our estimates. This is different if
the cut line is fermionic, in which case it must belong to the
$q$-$p$-trajectory. We find then a propagator singularity
$\big((q+L)^2-m^2\big)^{-1}$ with $L=\sum'(\pm\ell_i)$, $\sum'$
summing over a subset of $\sigma$.  This factor diverges like
$\Delta^{-1}$ and this divergence is in general not cancelled by
an additional factor of order $\Delta$. Moreover, even 1PI bubbles
are in general singular at the mass shell of their external
fermionic variables, due to IR effects. These IR singularities
are, however, not of power type, but only powers of logarithms.

We need to show that these internal singularities do not overpower
the vanishing of the phase space.    Notice that the left-hand
side of Eq.~\Ref{18} shown explicitly in Eq.~\Ref{20} contains the
factor $\delta_-(q)$ which is singular of order 1 in our way of
counting. Such a singularity must then also be present in the
right-hand side.  But stronger singularities are not admissible.
We need, then, to determine the strongest $\Delta\to0$
singularities that may occur in our graphs. In S-graphs this
maximal singularity is given by their ``web" parts defined in
Sect.\ 15.2 of BK, to which we refer for details\footnote{The
derivation of the web rules in BK contains gaps. But the result is
correct as stated.}. The replacement of the full graphs by their
webs commits an error of order $\Delta$, hence is justified for
our purposes. The same is true for the j-graphs containing the
 vertex of Fig.~3. But this situation is changed in
going over to the M-vertex of the M-graph defined by
Eq.~\Ref{28}, because the derivations in this expression may, and
occasionally will, produce a non-negligible singularity in the
correction term. We will return to this important point later on.
At the moment we consider only the web contributions.

A web graph in the rightmost sector $S_3$ consists of the part of
the trajectory (called a semitrajectory) contained in $S_3$, of a
vertex lying on it  for each $\ell_i\in L_r\cup L_m$, and of
internal photon lines with momenta $\{u_j\}$ connecting the
semitrajectory to itself. The $u_j$ are integrated over. The
trajectory propagators are, apart from numerical factors of no
present interest, of the form\footnote{Notice the presence of a
factor $\delta_-(s)$, meaning that $\Delta$ in the web rules of BK
(p.258) vanishes. The $\Delta$ of BK should not be confused with
our present $\Delta$ which is differently defined.}
$\big((s,L)-i\varepsilon\big)^{-1}$ with $L$ a non-empty sum over
$\ell_i\in \{L_r\cup L_m\}$. Since $s-q$ is negligible in the
static limit we can replace $s$ by $q$ in this expression. At each
vertex we have a factor $s^{\mu}$, replaceable by $q^{\mu}$. $\mu$
is the index occurring in the vertex factor $\gamma^{\mu}$ of the
original full graph from which the web is derived. Momentum is
conserved  at each vertex (the photon lines are directed from left
to right in Figs.\ 1 and 2). If no $u_j$-lines are present, then
there exist $|L_r|+|L_m|$ singular propagator factors  producing a
$\Delta^{-|L_r|-|L_m|}$-singularity in the static limit. The
presence of $u_j$-lines leads to additional IR singularities of
logarithmic type. $u_j$-lines can produce SEPs which at first
sight even increase the order of the power singularities. But this
effect is cancelled by the inclusion of appropriate mass
renormalization vertices in the web rules. The presence of a
physical $\Psi_n$-vertex at the end of the trajectory may
introduce $n$ further $\Delta$-singularities coming from the
factors $\tilde{r}(\ell_j)$ in the rules pertaining to physical
vertices. But this is compensated by  a corresponding lowering of
the number of trajectory propagators.  In the same way we find in
$S_1$  a singularity of order $\Delta^{-|L_{\ell}|-|L_m|}$
possibly multiplied by a weak IR singularity.

In the central $T^+$-sector $S_2$ the trajectory starts and ends
at a cross line. It is found that the web construction can be
started at either end resulting both times in the same singular
behavior. In a S-graph this yields a product of
$|L_{\ell}|+|L_r|-1$ singular factors of the form $(s,L)^{-1}$  if
there are no $u$-lines, hence a singularity of order
$\Delta^{-|L_{\ell}|-|L_r|+1}$. This remains true up to weak
singularities if $u$-lines are present.

   Hence we find in an S-graph a total power singularity of strength
 \[ \Delta^{-2(|L_{\ell}|+|L_r|+|L_m|)+1}\ ,  \]
  which together with the
phase space \Ref{33} gives a resultant singularity of order
$\Delta^{-1}$, possibly multiplied with a weak IR singularity.
Apart from the IR complications this is the desired behavior, as
has been remarked earlier.

In the central sector of a j-graph we have, besides the
$\ell$-vertices, also a current vertex with momentum $k$ as shown
in Fig.~3. It occurs in our final expression in the form \Ref{28}.
Hence we are only interested in values of ${\bf k}$ in an
arbitrarily small neighborhood of the origin, and $k_o=r'_o-s'_o$
vanishes in the static limit. The leading singularity of this
sector can be found exactly like in S-graphs, simply treating the
k-vertex like an additional $\ell$-vertex.
Because of the additional vertex on the web trajectory, the order
of the singularity is at first increased by 1 relative to the
corresponding S-graph. But, inserting this into \Ref{28} we see
that the $k_o$-integration lowers the singularity strength by one
order. On the other hand, the $k$-derivations tend to worsen the
singularity again. In order to see this we write the ${\bf
k}$-dependent part of the first term in the last form of \Ref{28}
as
 $ \frac{\der}{\der k_1}\,
\delta^3\big(\mathbf{k+s-r+\sum(\pm\ell_i})\big)$
 where $r$ and $s$
are the momenta of the cut trajectory lines.  We can then transfer
the $k$-derivations to the other factors of the integrand via
integration by parts. And then we can use the remaining $\delta^3$
 to integrate over ${\bf p}$, resulting in the replacement
$\mathbf{r\to s+k}$ in the other factors. Applying
$\frac{\der}{\der k_1}$ to the propagator
 \[ \left(s,\,{\sum}'(\pm\ell_i) + {{\sum}''}(\pm u_j) +k\right)^{-1} \]
gives
 \[ \frac{s^1}{(s,\,\sum' + {{\sum}''}+k)^2} \]
which raises the order of the singularity by 1. Here $\Sigma'$ and
${\Sigma''}$ are partial sums over $\ell$'s and $u$'s respectively. But associated
with this $k_1$-derivative is a vertex factor $s^2$, so that we
obtain in the numerator the symmetric factor $s^1s^2$. The term in
question is thus cancelled by an analogous term in the
$k_2$-derivation part. The singular factor $\delta_-(r)$ can be
removed by using it to first integrate over $r_o$ before
differentiating. There remains the factor $\tilf^*(p)$, which
depends on $\mathbf{p=s+k+\sum'\ell_i}$ only in the combination
$|{\bf p}|^2$. Hence
 \beq
  \frac{\der}{\der k_i}\,\tilf\Big|_{{\bf k}=0} = 2\big(s_1+{\sum}'\ell_{i,1}\big)
  \,\tilf'(|{\bf p}|^2)  \lb{A}  \eeq
where $f'$ is the derivative of $\tilf$ with respect to $|{\bf
p}|^2$ and the $p_o$-dependence of $\tilf$ has been ignored. The
bracket on the right-hand side is small in the static limit, but
this is not sufficient to make the term negligible, because the
static approximation relies on the assumption that the support of
$\tilf$ is tiny, which implies that $\tilf'$ is large. If $D$ is
the diameter of the support, then $\tilf'$ is large of order
$D^{-2}$ relative to $\tilf$ itself, while the factor
$(s_1+\cdots)$ is only small of order $D$. But there is also the
vertex factor $s^2$ of the k-vertex in the web which is again of
order $D$. Together, these two small factors still do not render
the $\tilf$-derivative negligible, but at least the term does not
explode for $D\to0$. And it is multiplied with a web of the
original, undifferentiated, j-type. This will turn out to be
important. Using these results we obtain for M-graphs the same
$\Delta^{-1}$ behavior, up to weak singularities, as for the
S-graphs.

We turn now to the problem of the possible relevance in M-graphs
of terms neglected in going over from the full graph to its web.
In BK it has only been shown that these terms are less singular
than the web by one order. Hence the $k$-derivatives might produce
singularities of the same order as that of the web, which could
not be neglected. The construction of the web as explained in BK
proceeds in several steps. In the first step it is shown that the
relevant singularity is correctly described by graphs not
containing any closed fermionic loops, but with their vertices and
propagators acquiring additional, but finite, numerical factors.
In this step it was used that  a loop integral vanishes at the
origin of its external variables. But this vanishing is actually
of second order, which  makes the $k$-derivation innocuous. In the
following steps of the construction the propagators and vertex
factors of the trajectories of the remaining graphs are simplified
by a procedure acting locally in an ``active region" which sets
out from the $s$-line (the cross line from $S_3$ to $S_2$) and
moves along the trajectory until it reaches the $r$-line. In the
part of the trajectory already traversed by the active region the
web rules hold, ahead of it the original Feynman rules. The active
region itself consists of a difference of the two forms (an
example will be shown presently). The ``static'' singularity of
the active region is better by at least one order than in either
the full or the web graph. We  transfer the $k$-derivations from
$\delta^3$ to the other factors and use then $\delta^3$  to
integrate over ${\bf r}$. The remaining independent variables are
$s,\,k,\,\ell_i,\,u_j$. The semitrajectory before the k-vertex
(that is on its $s$-side) is $k$-independent and therefore not
involved in the differentiations. The $k_1$-derivative of the web
propagator  $(s,L+k)^{-1}$ after the k-vertex, $L$ a sum of
$\ell_i$s and $u_j$s, is cancelled against a $k_2$-derivative as
explained earlier. The $k_1$-derivative of the full propagator
singularity $\big[(s+L+k)^2-m^2\big]^{-1}$  is $- (s^1+L^1+k^1)\,
\big[\cdots\big]^{-2}$.  The increased order of singularity is
offset by the vanishing of the numerator $(s^1+L^1)$ in the static
limit (remember $k^1=0$), so that the weakening of the singularity
from the active region remains effective. If the active region
lies before the k-vertex it gets not differentiated. If it lies
after the k-vertex, its differentiation may restore the dangerous
degree of singularity of the original graph. But in this case the
k-vertex belongs to the web and carries the small factor $s^1$ or
$s^2$, so that the term remains negligible in the static limit. The
critical case is that of the active region containing the
k-vertex. It has then essentially the form
 \[ (\sla{s}'+m)\,\gamma^i-2s^i = 2({s'}^i-s^i)- \gamma^i(\sla{s}' +\sla{k}-m)
+\gamma^i\sla{k}\ ,\quad i=1\ {\rm or}\ 2\ ,  \] where $s'$   is the
trajectory momentum entering the k-vertex. The factor
$(\sla{s}'+m)$ is the numerator of the $s'$-propagator. The
difference $({s'}^i-s^i)$ in this expression is $k$-independent
and vanishes in the static limit, hence is not causing problems.
The factor $(\sla{s}'+\sla{k}-m)$  multiplied into the next
propagator $(\sla{s}'+\sla{k}-m)^{-1}$   removes the latter's
singularity, giving the $k$-independent value 1. The factor
$\sla{k}$   in the last term is small, but its $k$-derivatives are
not. This term persists and must not be neglected. As is easily
seen    from \Ref{28} it produces a vertex factor
$[\gamma^2,\gamma^1]=-4i\Sigma_3$. As a result we obtain from the
correction terms a non-negligible contribution of the same form as
the j-web, except that in the k-vertex we have a vertex factor
$-4i\Sigma_3$ instead of $2s^i$. Notice that this $\Sigma_3$  is
the only surviving factor containing $\gamma$-matrices, so that it
commutes with the other web factors. This term occurs in addition
to the term found earlier, which contains an ordinary j-web
multiplied with a $\tilf$-derivative. Both these terms have the
correct power singularity but may in addition contain weak IR
singularities.

Finally we must note that an important problem concerning IR
divergent SEPs has been suppressed in our considerations. To wit:
the external  variable of a SEP next to a cross line, e.g. the
$s$-line, not being separated from it by a VP (= vertex part), is
restricted to the mass shell. Hence that SEP is in general
divergent, that means non-existent, not merely singular. But we
know from Chap.\ 11 of BK that these divergences cancel between
graphs with the same scaffolding\footnote{A scaffolding is a
Feynman graph not yet divided into sectors.}.   This means that
the said divergence in a given L-class (defined by the numbers
$|L_{\ell}|,\;|L_r|,\;|L_m|$) cancel against corresponding
divergences in other L-classes. And this cancellation happens
identically, not only in the static limit. Therefore the
dependence on the neglected ``small" external variables
$\Delta,\;\mathbf{p,\;q},\;k_0$, must be the same in all classes,
so that the divergences occurring in the separate classes are
irrelevant because they cancel in the sum over classes.
Alternatively we could circumvent this problem by not integrating
over the internal variables  $u_j$ at once, working at the level
of integrands instead of integrals, as was habitually done in BK.

 \medskip

We turn to proving the cancellation of the undesirable terms of
the right-hand side of Eq.~\Ref{18} containing the combinations
$\tilf$--$\tilf^*$, $\tilf_M$--$\tilf^*$, $\tilf$--$\tilf^*_M$. We
start with the case $\tilf_M(q)$--$\tilf^*(p)$. This combination
occurs in the M-graphs with $|L_{\ell}|>0$, $|L_r|=|L_m|=0$, but
{\em not} in the corresponding S-graphs, because there we have
$p=q$ so that if $q$ is restricted to the mass shell, so is $p$.

It has been shown above that the leading $\Delta$-singularity is
of the same order $1/\Delta$  as that of the left-hand side of
Eq.~\Ref{18}, and that it can be expressed as a sum of two terms,
both of j-web form. The first contains a $\tilf$-derivative, the
second a factor $\Sigma_3$ at the k-vertex. Hence the contribution
of a single M-graph is not negligible. However, if we permute the
vertices of a given web in the central sector, we obtain another
legitimate web. And summation over these permutations removes the
leading singularity, as will now be shown. A trajectory propagator
of the central web is of the form (up to irrelevant constant
factors) $\big(q,\,L'+U'(+k)\big)^{-1}$, where $L'$ and $U'$ are
partial sums of cross variables $\pm\ell_i$ and internal photon
variables $\pm u_j$\footnote{$u_j$ is IR critical
only at $u_j=0$.} respectively, and the term $+k$ may or may not
be present. We have started the web construction at the $q$-end.
Summing the product of these factors over all permutations of the
vertices yields (see BK, p.326)
 \beq
 \prod_j\left(-\, \frac{1}{(q,u_j)^2}\right)\,\prod_i\,\left(
 \frac{1}{(q,\ell_i)}\right)\,\frac{1}{(q,k)}\, \big(q,\,L+k\big)
 \lb{34}           \eeq
with $L=\sum_{L_{\ell}}\ell_i$. There is one factor more in the
denominator (notice that the $u$-factors are integrated over and
are thus irrelevant for the power behavior) than in our previous
estimate of the singularity strength. But we also have the
additional factor $(q,\,L+k)=\big(q_o(L^o+k^o)
-(\mathbf{q,\,L+k})\big)$ which multiplies the cross propagators
$\delta\big(q^o+L^o+k^o+\omega(\mathbf{q+L+k})\big)\,
\delta\big(q_o+\omega({\bf q})\big)$. This implies that
 \[ L^o+k^o = - \omega(\mathbf{q+L+k}) + \omega(\mathbf{q}) \]
which vanishes of second order in the static limit ${\bf q}\to
0,\;{\bf L}\to 0$, remembering that ${\bf k}=0$. This is then also
true for $(q,L+k)$. Hence the expression \Ref{34}  vanishes
stronger by one order than the individual terms of the sum, and
this suffices to make the $\tilf_M$--$\tilf^*$ term vanish in the
static limit. Note that this argument works only if at least one
$\ell$-line is present.

The vanishing of the $\tilf(q)$--$\tilf^*_M(p)$ term, that is the
term with $|L_{\ell}|=|L_m|=0,\;|L_r|>0$, is shown in the same
way.

$\tilf(q)$--$\tilf^*(p)$ terms occur in graphs in which neither of
the external momenta is restricted to the mass shell by a factor
$\delta_-$, that is in the graphs in which both $|L_{\ell}|+|L_m|$
and $|L_r|+|L_m|$ are positive. The corresponding M-graphs with
$|L_r|>0$ (or similarly with $|L_{\ell}|>0$) can be shown to
vanish in the static limit in the same way as in the
$\tilf_M$--$\tilf^*$ case. Just replace $q$ by the cross momentum
$s$ and $\ell_i$ by $-\ell_i$ for photon lines crossing into the
$q$-sector. But note that this result relies on cancellations
between graphs with permuted vertices in the central sector. Hence
it does not apply to the case $|L_{\ell}|+|L_r|=0$.

For S-graphs with $|L_{\ell}|+|L_r|>1$ we can again use the same
method to show their irrelevance. The S-graphs with
$|L_{\ell}|+|L_r|=1$ vanish if defined as limits $\mu\to0$ from
massive QED as mentioned near the beginning of this section. For
instance, if $|L_{\ell}|=1,\;|L_r|=0$, then the two fermionic
cross momenta $r,\;s$, of Fig.~1 are related by $r=\ell+s$ with
$\ell$ the momentum of the only $L_{\ell}$-line. But this relation
cannot be satisfied if $\ell^2=\mu^2>0$ because both $r$ and $s$
are restricted to the negative mass shell. So, like in the M-case,
only the graphs with $|L_{\ell}|=|L_r|=0$ remain. Remember that
then we must have $|L_m|>0$ in order to get a $\tilf$--$\tilf^*$
term.


\begin{figure}[ht]
\begin{center}
\begin{picture}(310,110)
\SetOffset(0,-7)
 \SetWidth{0.8}
 \DashCArc(155,-102)(212,60,120){3}
 \DashCArc(155,-112)(212,60,120){3}
 \ArrowLine(20,70)(0,70)   \Line(70,70)(80,70)
 \Line(110,70)(130,70)  \Line(180,70)(200,70)
 \Line(230,70)(240,70)  \ArrowLine(310,70)(290,70)
 \Text(10,73)[b]{$p$}  \Text(300,73)[b]{$q$}
 \SetWidth{0.5}
 \GOval(45,70)(15,25)(0){0.6}
 \GOval(265,70)(15,25)(0){0.6}
 \GOval(155,70)(15,25)(0){0.6}
 \GCirc(95,70){15}{0.6}  \GCirc(215,70){15}{0.6}
 \Text(45,70)[]{$B_1$}  \Text(95,70)[]{$B_2$}  \Text(155,70)[]{$B_3$}
 \Text(215,70)[]{$B_4$}  \Text(265,70)[]{$B_5$}
 \SetWidth{0.3}
 \Line(120,117)(120,50)  \Line(190,117)(190,50)
 \SetWidth{0.5}
 \Line(75,44)(235,44)    \Line(75,44)(75,52)
 \Line(235,44)(235,52)
 \Text(155,40)[t]{Subgraph ${\cal S}$}
 \Text(155,7)[b]{{\bf Fig.~4.} \small $\tilde{f}$--$\tilf^*$ graph}

\end{picture}
\end{center}
\end{figure}

The graphs in question are of the form shown in Fig.~4.
We consider now the full graphs, not webs. $B_2$ and $B_4$ are
chains of SEPs. $B_3$ is a chain of SEPs in an S-graph, and such a
chain interrupted at one place by a $M_3$ vertex part in an
M-graph. $B_1$ and $B_5$ are chains of SEPs, and 1PI graphs
connected to at least one $\ell$-line, such that the link next to
$B_2$ or $B_4$ respectively is of the latter type (i.e. not a
SEP). We fix the perturbative order $\sigma$ and $B_1$, $B_2$, but
sum over all possible subgraphs $\mathcal{S}$ of the appropriate
order $\rho<\sigma$ occurring in any of the terms of the
right-hand side of Eq.~\Ref{18}, including their numerical factors,
in particular the factors $g_{\tau}$. We find that these $\mathcal
{S}$-terms are simply the $\tilf_M$--$\tilf^*_M$ terms of
\Ref{18} taken in order $\rho$, and that they sum to zero if
this equation is assumed to be satisfied to all orders lower than
$\sigma$.

\medskip

We are now in a position to solve Eq.~\Ref{18} for the unknown
$g_{\sigma}$. The left-hand side $L$ is well defined and depends
only on $\tilf_M$. Explicitly it is given by \Ref{20} with
$g_o$ replaced by $g_{\sigma}$:
 \beq
 L = \frac{g_{\sigma}}{4m}  \int
 \frac{d^3q}{2\omega}\,\tilf(-\omega,\mathbf{q})\,(\omega\gamma^o-q_i\gamma^i-m)\,
 \gamma^o\,\tilf^*(-\omega,\mathbf{q})\ ,     \lb{35}
 \eeq
 $\omega=\omega(\mathbf{q})$. We must show that the surviving
 $\tilf_M$--$\tilf^*_M$ term in the right-hand side exists, is
 well defined, and is of the form \Ref{35} up to a numerical
 factor. The contributing M- and S-graphs are those with $|L_{\ell}| = |L_r| = |L_m| =
 0$. They are of the form shown in Fig.~4, but without the extremal
 bubbles $B_1$ and $B_5$. Remember that $B_2$ and $B_4$ are
 (possibly empty) chains of SEPs. We show first by general
 induction that the S-graphs and M-graphs cancel unless $B_2$ and
 $B_4$ are empty. Assume this to be true in lower orders. Keep
 $B_{2,4}$ fixed and sum over all $B_3$ of the relevant order in
 all terms of Eq.\Ref{18}. These terms are exactly  the surviving
 terms of \Ref{18} in this lower order, hence they cancel in the
 static limit if the problem has already been solved in this
 order.

 Concerning the existence of the remaining graphs we are faced
 with two problems. The first problem is that of the ``collinear
 singularities" which are present even in massive QED. In this
 theory, a SEP in the central sector with external momentum $q$
 is, after mass renormalization, of the form $(q^2-m^2)\,\Sigma(q)$
 with $\Sigma$ continuous at the mass shell $q^2=m^2$. In an
 M-graph we find then on the $q$-side of the $M_3$ vertex-part a
 product
 \beq
 \left(\frac{1}{Q+i\varepsilon}\,(Q+i\varepsilon)\,\Sigma(q)\right)^{\!\alpha}
 \theta(-q_o)\,\delta(Q)        \lb{35a}
 \eeq
 where $Q=q^2-m^2$ and $\alpha$ is the number of SEPs present.
 This expression is at first undefined because the product $(Q+i\varepsilon)^{-1}\, Q\,
 \delta(Q)$ is not associative and therefore ill defined. We solve
 this problem by defining
 \beq
 \delta(Q) = \frac{i}{2\pi}\,\left(\frac{1}{Q+i\varepsilon}- \frac{1}{Q-i\varepsilon}
 \right)                \lb{36}
 \eeq
 with the $\varepsilon$ occurring already in \Ref{35a}. This means
 that the limit $\varepsilon\to0$ must be taken simultaneously in
 all factors. This prescription is justified as follows. Using the
 Fourier transform
 \[ \tilde{\Delta}_{\pm}(q) = \pm i(2\pi)^{-3}\,\theta(\pm q_o)\,
 \delta(q^2-m^2) \]
 of the familiar singular functions $\Delta_{\pm}(\xi)$,
 Eq.~\Ref{36} is obtained by Fourier transform from the
 relation
 \[ \Delta_+(\xi)-\Delta_-(\xi) = \Delta_F(\xi) -\overline{\Delta_F(-\xi)} \]
 which is a basic ingredient in many calculations used in our
 formalism. For instance, the proof of the important Lemma 9.2 of
 BK, transposed into momentum space, works as stated in $x$-space
 only with the convention introduced here. Notice that
 \[ (Q+i\varepsilon)\,(Q-i\varepsilon)^{-n} = 1 +
 2i\varepsilon(Q-i\varepsilon)^{-n} = 1 \]
 in the sense of distributions, where the limit $\varepsilon\to0$
 is understood. With this definition the expression \Ref{35a}
 becomes $\theta(-q_0)\,\big((\Sigma(q)\big)^{\alpha} \delta(Q)$,
 which is well defined for $\mu>0$ ($\mu$ the photon mass). The
 same consideration applies of course to the $p$-side of the
 trajectory from the $M_3$ vertex part. For the S-graphs we
 distinguish two cases. If no SEP is present in the central
 section, then the trajectory is a single line leading directly
 from the external sector $S_3$ to the external sector $S_1$, and
 by our rules the corresponding propagator is the well defined
 expression $(\sla{q}+m)\,\delta_-(q)$. If there are
 $\alpha>0$ SEPs present we find the ambiguous product
 \[ \theta(-q_o)\,\delta(Q)\,\big(\Sigma(q)\,(Q+i\varepsilon)
 \big)^{\alpha}\,(Q+i\varepsilon)^{-\alpha+1}\,\delta(Q)\ . \]
 This is again uniquely fixed by the definition \Ref{36} to be
 \[ \frac{i}{2\pi}\,\theta(-q_o)\,\big(\Sigma(q)\big)^{\alpha}\,
 \delta(Q) \]

 The second problem is the IR problem. $\Sigma(q)$ and the
 $k$-vertex part, even of a j-graph, diverge for $\mu\to0$ weakly
 (i.e. like a power of $\log Q$) at the mass shell, so that the
 expressions obtained above for individual graphs no longer exist.
 We must show that these IR divergences cancel between graphs  in
 the static limit. For this we need an explicit expression for
 $g_{\sigma}$. Let $m_{\tau}(q,p,\mathbf{k})$ be the sum over all
 3-line $T^+$-graphs of order $2\tau+1$ intersecting the
 trajectory and containing the $k$-vertex, before setting  ${\bf
 k}=0$. $q$ is the momentum of the entering fermion line, $p$ of the
 leaving one. Let $s_{\tau}(q)$ be the sum over all properly
 renormalized 2-line $T^+$-graphs of order $2\tau$ intersecting
 the trajectory. Let $m'_{\tau}$ and $s'_{\tau}$ be the analogous
 sums over 1PI graphs only. This means that $s'_{\tau}$  is a sum
 over SEPs, $m'_{\tau}$ a sum over VPs. Eq.~\Ref{18} becomes
 \beqa
   & & \frac{g_{2\sigma}}{4m} \int dq\,\delta_-(q)\,\tilf_M({\bf
   q})\,(\sla{q}+m)\,\gamma^o\tilf^*_M({\bf q})  \nonu
    \gl -\int dp\,dq\,\tilf_M({\bf q})\,\delta_-(q)\,(\sla{q}+m)\,
 m_{\sigma}(q,p,{\bf k})\, (\sla{p}+m)\,\gamma^o\delta_-(p)\,
 \tilf^*_M({\bf p})\Big|_{{\bf k}=0}      \nonu
   & & \hspace*{-1cm} -\frac{1}{4m} \sum^{\sigma-1}_{\tau=0} g_{2\tau}\int dq\,
   \delta_-(q)\,\tilf_M({\bf q})\,(\sla{q}+m)\,
   s_{\sigma-\tau}(q)\,(\sla{q}+m)\,\gamma^o\delta_-(q)\,
   \tilf^*_M({\bf q})\ ,   \lb{37}
 \eeqa
 both sides to be taken in the static limit. This expression is at
 first purely formal, since the individual terms on the right do
 in general not exist. The following operations, which are largely
 algebraic, will at first also be carried out at this formal
 level. They can be given a strict meaning by arguing at the level
 of (properly subtracted) integrands before carrying out the
 integrations over internal momenta.

 In the S-terms on the right-hand side we use that after mass
 renormalization $s_{\rho}$ is, for $\rho>0$, of the form
 \[ s_{\rho}(q) = -2\pi i\,(\sla{q}-m)\,T_{\rho}(q) \]
 where the spin scalar $T_{\rho}$ is finite at $q^2=m^2$ in the
 massive case $\mu>0$, while it develops there a weak singularity
 for $\mu\to0$. Using Eq.~\Ref{36} and noticing that the factor
 $(q^2-m^2)$  must be read as $(q^2-m^2+i\varepsilon)$ in a
 $T^+$-sector we find that
 \beq
 -2\pi i\,\delta_-(q)\,(q^2-m^2)\,T_{\sigma-\tau}(q)\, (\sla{q}+m)
 \,\gamma^o\,\delta_-(q) = T_{\sigma-\tau}(q)\,(\sla{q}+m)\,
 \gamma^o\delta_-(q) .    \lb{38}
 \eeq
 The $q_o$-integration can be carried out with the help of the
 factor $\delta_-$, producing a new factor
 $(2\omega_{\mathbf{q}})^{-1}$ and the replacement of $q^o$ by
 $-\omega_{\mathbf{q}}$ in the remaining factors. This meaning of
 $q^o$ will be understood in the sequel.

 The M-term in Eq.~\Ref{37} can be written as
 \beqa
 -\sum_{\alpha+\beta+\gamma=\sigma}
  \int dp\:dq\,\delta_-(q) \,\delta_-(p)\,
 \tilf_M(q)\,(\sla{q}+m)\, T_{\alpha}(q) \,m'_{\beta}(q,p,{\bf
 k})& &  \nonu
 & & \hspace{-9cm} \times\ T_{\gamma}(p)\,(\sla{p}+m)\,\gamma^o\tilf^*_M({\bf p})
 \Big|_{{\bf k}=0} \  .  \lb{39}
 \eeqa
 We have defined $T_o=1$. The factors $\delta_-$ can be removed by
 integration over $p_o$, $q_o$, resulting in a factor $(4\omega_{{\bf q}}
 \omega_{{\bf p}})^{-1}$ and the replacement of $q_o$, $p_o$, by
 $-\omega_{{\bf q}}$, $-\omega_{{\bf p}}$, respectively. It seems
 at first troublesome hat the factor $\tilf^*$ depends  on
 $\mathbf{p}$ instead of $\mathbf{q}$ like in the S-terms. This
 problem is solved as follows. We remember that the relevant
 contribution to the integrand of \Ref{39} is a sum of two terms.
 The first term consists of a j-web multiplied with a
 $\tilf$-derivative of the form \Ref{A}, but without any
 $\ell_i$'s. That is, the factor in front of $\tilf'$ is simply
 $s_1$ or $s_2$, and it is multiplied with a factor $s_2$ or $s_1$
 at the k-vertex of the j-web, which leads to the familiar
 cancellation between the two terms in the definition of $M_3$.
 This term does therefore {\em not} contribute! The remaining
 second term contains a web which is a j-web except that the
 $k$-vertex carries a factor $\Sigma_3$. No derivatives are present
 in this term, hence we can set ${\bf k}=0$ in
 $\tilf^*_M(\mathbf{q+k})$ without further ado, obtaining the
 desired result $\tilf^*_M({\bf q})$. Therefore we can replace in
 the expression \Ref{39} the argument ${\bf p}$ of $\tilf^*_M$ by ${\bf
 q}$ without changing the result. The relevant part of
 $m'_{\beta}$ can be written as $-(2m)^{-1}\Sigma_3U_{\beta}(\mathbf{q,p})\,
 \delta^3(\mathbf{k+q-p})$, $U_{\beta}$ a spin scalar. And it can be
 treated as an isolated factor, like $T_{\gamma}$, not containing
 differential operators possibly acting on the other factors. This allows us to
 take the static limit termwise. Moreover, the relevant web contributions to
 $T$ are spin scalars, while $m'_{\beta}$ contains $\Sigma_3$ as the only
 spinorial factor. And
 $\Sigma_3$ commutes with $\gamma^o$, hence with $(\sla{q}+m)$ in
 the static limit.

Using $\Sigma_3\tilf_M = \half\,\tilf_M$ and dropping in the
static limit of Eq.~\Ref{37} the factor
\[ (8m)^{-1}\int d^3q\,\tilf_M({\bf q})\,(\gamma^o-1)\,\tilf^*_M({\bf q}) \]
which is common to all the terms, we find
 \beq
g_{2\sigma}   =\bigg[\sum_{\alpha+\beta+\gamma=\sigma}
T_{\alpha}(q)\,U_{\beta}(q,p)\,T_{\gamma}(p) -
\sum_{\tau=0}^{\sigma-1}g_{2\tau}T_{\sigma-\tau}(q)\bigg]_{\mathbf{p=q}}
\lb{41}\ .   \eeq
  Here the limit
 $q=p\to(-m,\vec{0})$ must be taken, whose existence has not yet
 been established. At the moment we consider therefore \Ref{41} as
 defining a function $g_{2\sigma}(q)$ defined in a small
 neighborhood of the mass shell. We use the following lemma:

\medskip

{\bf Lemma.}
 \begin{it}
 $g_{2\sigma}$ as defined by Eq.\Ref{41} can be written as
\beq
 g_{2\sigma} = \sum_{\tau=0}^{\sigma}U_{\tau}(q,p)\, T_{\sigma-\tau}(p)
 \big|_{p=q}\ .    \lb{42}
\eeq
 \end{it}
 The proof proceeds by induction with respect to $\sigma$. The result is clearly correct
 for $\sigma=0$ (remember $T_o=1$). Let $\sigma>0$ and assume that \Ref{42} holds for
$g_{2\tau}$ with $\tau<\sigma$. We insert this inductive ansatz
into \Ref{41} and obtain, using that $T_{\alpha}$ and
$U_{\beta}$ commute,
 \begin{eqnarray*}
 g_{2\sigma} \gl
 \big(\sum_{\alpha+\beta+\gamma=\sigma}T_{\alpha}(q)\,U_{\beta}(q,p)
\,T_{\gamma}(p)   \\
 & &  -\sum_{\alpha+\beta+\gamma=\sigma}T_{\alpha}(q)\,
 U_{\beta}(q,p)\,T_{\gamma}(p)
 +\sum_{\tau=0}^{\sigma}U_{\tau}(q,p)\,T_{\sigma-\tau}(p)\big)
 \big|_{p=q}\ ,
 \end{eqnarray*}
  the claimed result.

 \medskip

The expression \Ref{42} is IR finite. We know that the possible IR
divergences are given by replacing $U_{\tau}$ and
$T_{\sigma-\tau}$ by their webs. The web construction starts from
the $q$-end of the trajectory in the central sector $S_2$, and $q$
is restricted to the mass shell. The webs in question contain only
one vertex associated with an external variable, namely the
$k$-vertex. The photon lines incident at the other vertices are
internal lines of $S_2$, starting and ending at the trajectory. It
follows from the proof of Lemma 17.1 in BK that in the sum over
all webs of this form only the graphs survive in which no vertices
exist {\em after} the $k$-vertex. This also excludes photon lines
starting after the $k$-vertex  and ending in front of it. And SEPs
situated in front of the $k$-vertex are not present because such
terms are not present in the expression \Ref{42}.

\section{Equality to the Conventional Result}

In this final section it will be shown that our result \Ref{42}
agrees with the conventional result. We can avoid the IR problem
by starting from massive QED and then taking the limit $\mu\to0$.
This is legitimate in our method, as has been shown in the
preceding section. On the other hand, we have argued that this
procedure is of doubtful legitimacy in the conventional approach.
But it {\em is} the method used there, so that the comparison of
the two results rightfully employs it.

The conventional approach finds the magnetic moment of the
electron from an investigation of its scattering in an external
magnetic field (see, e.g., \cite{Kin}, \cite{Sak}, \cite{Wei}).
The corresponding scattering amplitude is essentially given by the
3-point Green's function $\Gamma^{\mu}(q,p)$, where $q$ is the
momentum of the incoming electron, $p$ that of the outgoing one,
and $k=p-q$ is a photon variable not explicitly shown. The
external $k$-vertex is a current vertex with the
vertex factor $(2\pi)^{-3/2}\gamma^{\mu}$. $\Gamma^{\mu}$ occurs
in the amplitude in the sandwiched combination
 \beq
\delta_-(q)\,(\sla{q}+m)\,\Gamma^{\mu}(q,p)\,(\sla{p}+m)\,\delta_-(p)\
.      \lb{43}        \eeq

The electron field is conventionally renormalized such that its
clothed propagator has the same pole at the mass shell with the
same residue as the bare propagator.\footnote{Remember that this
is not possible in the case $\mu=0$ because of the IR problem.}
Graphs containing SEPs in the variables $q$ or $p$ do therefore
not contribute to the expression \Ref{43}, so that $\Gamma_{\mu}$
may be replaced by the proper vertex part $\Lambda^{\mu}$ defined
as a sum over 1PI graphs only. From covariance and the
Ward-Takahashi identities it follows\footnote{For a proof see
e.g. Sect.10.6 of \cite{Wei}} that $\Lambda^{\mu}$, sandwiched
like in \Ref{43}, can be decomposed as
 \beq
\Lambda^{\mu}(q,p) = \frac{e}{(2\pi)^{3/2}}\Big[\gamma^{\mu}
F_1(k^2) + \frac{1}{2m}\,k_{\nu}\sigma^{\mu\nu} F_2(k^2)\Big]\ .
\lb{44}   \eeq
 $F_1$ and $F_2$ are the functions occurring in Eq.~(1.5) of
 Ref. \cite{Kin}. The coefficients in the two expressions differ due
 to the use of different conventions. Charge renormalization, that
 is the condition that the coupling constant $e$ is equal to the
 charge of the positron, demands that $F_1(0)=1$, so that
 $F_{1,\sigma}(0)=0$ for $\sigma>0$, when $F_{i,\sigma}$ is the
 coefficient of $e^{\sigma}$ in the perturbation expansion of
 $F_i$. Moreover, $F_{1,\sigma}$ is $\mathcal{C}^{\infty}$ at
 $k=0$ if $\mu>0$. For $\mu=0$ its strongest possible singularity
 at $k=0$ is of the form $k^2(\log k^2)^n$, $n$ a positive
 integer. Hence
  \[ \frac{\der}{\der k_{\mu}}\,F_{1,\sigma}\Big|_{k=0}=0\ . \]
For the gyromagnetic ratio one finds
 \beq
g_{\sigma} = 2\,F_{2,\sigma}(0)      \lb{45}
 \eeq
for $\sigma>0$. Remember that the equivalence of the two methods
in the order $\sigma=0$ has already been established in Sect.\ 3.

According to the results just mentioned, $F_{2,\sigma}(0)$ may be
determined from
 \beq
\frac{\der}{\der k_{\nu}}\,\Lambda^{\mu}_{\sigma+1} \Big|_{k=0} =
\frac{-i}{2m(2\pi)^{3/2}}\,\sigma^{\mu\nu} F_{2,\sigma}(0)
 \lb{46}  \eeq
or, more particularly, from
 \beqa
\frac{i}{2}\,\left(\frac{\der}{\der k_1}\,\Lambda^2_{\sigma+1} -
\frac{\der}{\der
k_2}\,\Lambda^1_{\sigma+1}\right)_{p=q=P} \gl
\frac{1}{m(2\pi)^{3/2}}\, \Sigma_3 F_{2,\sigma}(0)  \nonu
 \gl \frac{1}{2m(2\pi)^{3/2}}\,F_{2,\sigma}(0)   \lb{47}
\eeqa
 if $\Sigma_3=1/2$.

In the method proposed in this paper we have on the one hand
 \begin{eqnarray*}
  m'_{\tau} \gl -\,\frac{i}{2}\,(2\pi)^{3/2}\, \Big(\frac{\der}{\der k_1}\,
 \Lambda^2_{2\tau+1} - \frac{\der}{\der
 k_2}\,\Lambda^1_{2\tau+1}\Big)_{p=q=P}  \\
  \gl -\,\frac{1}{2m}\,F_{2,2\tau}(0)\ ,
 \end{eqnarray*}
on the other hand
 \[ m'_{\tau} = -(4m)^{-1}\,U_{\tau}\ , \]
where again $\Sigma_3=1/2$ has been assumed. Hence
\beq
 U_{\tau} = 2\,F_{2,2\tau}(0)\ .    \lb{48}
\eeq
 Inserting this into our result \Ref{42} we find agreement with
 the conventional result \Ref{45}. The $T$-containing terms
 $\tau<\sigma$ in \Ref{42}, which are not present in \Ref{45},
 correct for the fact that we use ``intermediate renormalization"
 instead of the conventional $\psi$-renormalization mentioned
 above. In intermediate renormalization the fermionic SEPs are
 subtracted at the origin instead of at the mass shell, thus
 avoiding the problem of the IR divergence of the renormalization
 constant $Z_2$. The proper position of the 1-particle singularity
 of the $\psi$-propagator, which is {\em not} a pole if $\mu=0$,
 is then secured by a finite mass renormalization. Of course, in
 the conventional method we have $T(p)=0$ at the mass shell, so
 that the $\tau<\sigma$ terms in \Ref{42} do not occur.

  \medskip

As a last remark we note that this equivalence proof tells us that
the UV finiteness of our $g_{\sigma}$ follows from the known UV
finiteness of the conventional result. There is no need to
renormalize (that is, subtract) our expression for $M_3$. This is
also seen by realizing that our result depends only on first
derivatives of the vertex part $\Lambda^{\mu}$, not on
$\Lambda^{\mu}$  itself. They are UV convergent since
$\Lambda_{\mu}$ is only logarithmically divergent.

\end{document}